# A Wall Function for Turbulent Boundary Layers under Rotation via Symbolic Regression

Yao Ma
*Research Institute of Aero-Engine*
*Beihang University*
Beijing, China
upwind@buaa.edu.com

Zhi Tao
*Research Institute of Aero-Engine*
*Beihang University*
Beijing, China
quick@buaa.edu.cn

Ruquan You*
*Research Institute of Aero-Engine*
*Beihang University*
Beijing, China
youruquan10353@buaa.edu.cn
*Corresponding author

Haiwang Li
*Research Institute of Aero-Engine*
*Beihang University*
Beijing, China
09620@buaa.edu.cn

***Abstract*—This study employs symbolic regression to derive physically interpretable, white-box wall-function expressions for turbulent boundary layers under system rotation. Flows in a rotating frame are subject to Coriolis forces, which deflect the boundary layer profile from static case. The classical law of the wall, formulated under non-rotating conditions, is ill-suited to describing the effects of rotation. To obtain the wall function under rotation, we examine the deflection behavior of the turbulent boundary layers on the leading and trailing sides, and construct wall functions that are valid over a wide range of rotation numbers. The analytical expressions show that, as the rotation effect intensifies, the boundary layer on the leading side contracts whereas that on the trailing side expands, and the leading side exhibits a tendency towards relaminarization, consistent with high-fidelity numerical results. The resulting symbolic expressions are compact and interpretable. The wall functions obtained in this study complement conventional wall functions, and provide a new avenue for turbulence model closure subject to system rotation.**

## I. Introduction

Rotating wall-bounded turbulence is widely encountered in nature and fluid machine, notably turbomachines. When the flow is observed in a rotating frame of reference, additional body forces, such as Coriolis force, the centrifugal force and the buoyancy force, are induced. The flow and heat-transfer characteristics deviate from those in stationary configurations. The flow resistance coefficient and convective heat-transfer coefficient in the internal cooling channels of rotating turbine blades in aero-engines can differ by as much as 50% from those under non-rotating conditions [1]. To investigate the influence of rotation on flow and heat transfer, and to elucidate the underlying physics of rotating turbulent flows, studies have largely relied on experiments and numerical simulations. Experimental facilities are, however, difficult to set up and prohibitively expensive, which restricts the deeper research of complicated flows [2-8]. Numerical simulations relying on turbulence models are ubiquitous in fluid mechanics, with the RANS (Reynolds averaged Navier Stokes) [9-16]. Yet RANS-based closures are usually calibrated under non-rotating conditions and contain empirical parameters that introduce non-physical unreality. Compounded by the non-conservative of the Coriolis force and the nonlinearity of the Navier–Stokes equations, conventional RANS models perform poorly in predicting flow and heat transfer under rotation, and discrepancies with experimental measurements can be as much as 50% [15].

The difficulty of predicting turbulent flow lies in the nature of turbulence as a multi-scale self-organizing system. The wall-bounded internal flow exhibits a series of characteristic scales. How rotation alters the dynamics at each of these scales remains an open question to be solved. If an appropriate non-dimensional characteristic quantities that collapse onto a universal form, the turbulent structures are said to exhibit similarity at that scale. Using the viscous length scale $\nu/u_\tau$ and the friction velocity $u_\tau$ yields the classical law of the wall. For example, we can get the viscous sublayer $u^+ = y^+$ and the logarithmic region $u^+ = 2.04 \ln y^+ + 5.5$. These universal formulas indicate that the viscous sublayer and the logarithmic layer are closely related to the near-wall activities. Motivated by this idea, Coles considered similarity laws that hold at larger scales and proposed a unified velocity profile spanning the logarithmic region through to the outer layer.

$$u^+ = \frac{1}{\kappa}\ln y^+ + B + \Pi\left(\frac{y}{H}\right) \tag{1}$$

Where κ and $B$ are universal constants, and $H$ is a global integral scale such as the channel half-width or the boundary-layer thickness. The wake function Π accounts for the influence of large-scale structures that are independent of viscosity. Furthermore, Spalding proposed an implicit wall function that encompasses the viscous sublayer, the buffer region, the logarithmic region and the outer layer:

$$y^+ = u^+ + e^{-\kappa B}\left[e^{\kappa u^+} - \sum_{n=0}^{3}\frac{(\kappa u^+)^n}{n!}\right] \tag{2}$$

Under rotating conditions, the turbulent boundary layer keeps the same form as its non-rotating counterpart: a viscous sublayer, a buffer region and a logarithmic region are still identifiable, but the shape parameters that characterize these regions vary with the rotation number, seen in Fig. 4 and Fig. 5. Consequently, some investigators have introduced rotation-dependent corrections to the coefficients of the law of the wall and obtained distributions that agree with reference data [2-8]. Such corrections, however, lack a clear physical justification. The present study therefore seeks to probe the mechanism through which system rotation modifies the boundary layer, with the aim of uncovering a simpler underlying physical description.

To overcome the poor predictive performance of turbulence models in rotating flows, numerous modifications have been proposed [11,12,17,18]. These efforts are categorized into two types. The first adjusts the empirical coefficients embedded in the models so that those match refernce data better; the second alters the expression form of the models, for example by adding or removing terms to improve accuracy. The first approach is straightforward and relies entirely on statistical fitting, with a sufficiently large sample it can, in principle, achieve arbitrary accuracy. But it offers no physical insight. The second approach may introduce physical corrections, yet constructing them is difficult and the terms are often fortuitous rather than systematic. To overcome and balance the shortcomings of both approaches, investigators have combined the power of statistical data-assimilation with the transparency and interpretability of symbolic manipulation, giving rise to symbolic regression — an optimization algorithm that employs symbolic operations to uncover the physical laws underlying the observed phenomena. This methodology has already been applied with success to aerodynamic design of aircraft [19-21].

This paper employs symbolic regression (SR) to combine the strengths of high-fidelity numerical data with the physical transparency of white-box models. Using heuristic genetic algorithms, we construct analytical wall functions for turbulent boundary layers that explicitly account for the effects of system rotation. This approach deepens the physical understanding of rotating wall-bounded turbulence, improves the predictive accuracy of turbulence models for system rotating flows, and advances the design of internal cooling in rotating turbine blades.

To highlight the role of the Coriolis force in system rotating flows, we take the internal flow in rotating turbine blade as a canonical configuration and idealize it as an incompressible, fully developed, spanwise-rotating plane channel flow bounded by two opposing walls. The mean velocity and Reynolds stress fields serve as the bridge connecting the rotational forcing to the resulting changes in the flow. Our objective is to reconstruct the velocity field under rotating conditions. The remainder of the paper is organized as follows. Section II describes the numerical approach employed in this study. Section III presents the typical results of the flow under rotation and the approach taken to find the wall function under rotation. Conclusions and an outlook are given in Section IV.

## II. Numerical Approch

### A. Governing Equations

We denote the velocity field by $\boldsymbol{u} = (u, v, w)^{\mathrm{T}}$, the pressure by $p$, the density by $\rho = 1$, the time by $t$, and the kinematic viscosity by $\nu$. Taking $x$ as the streamwise direction, $y$ as the wall-normal direction, and $z$ as the spanwise direction of the channel, the governing equations are:

$$\boldsymbol{\nabla} \cdot \boldsymbol{u} = 0 \tag{3}$$

$$\partial_t \boldsymbol{u} + (\boldsymbol{u} \cdot \boldsymbol{\nabla})\boldsymbol{u} = -\boldsymbol{\nabla} p + \nu \nabla^2 \boldsymbol{u} + 2\Omega \boldsymbol{k} \times \boldsymbol{u} \tag{4}$$

The dimensionless parameters are the friction Reynolds number $Re_\tau = u_\tau h/\nu$ and the rotation number $Ro_\tau = 2\Omega h/u_\tau$, where $\Omega$ is the angular velocity in $z$ direction and $\boldsymbol{k}$ is the unit vector in $z$ direction. In what follows, the wall towards which the Coriolis force acts on the main flow is referred to as the trailing side (near $y = 0$), and the opposite wall, towards which the Coriolis force is directed away from the bulk, is referred to as the leading side (near $y = 2h$). Here, $h$ is half the distance between the two parallel walls and $u_\tau = \sqrt{\tau_w} = \sqrt{\nu(\partial\overline{u}/\partial y)_w}$ is the friction velocity based on the wall shear stress. To obtain statistically stationary solutions, (1) and (2) are Reynolds-averaged. The Reynolds averaging employed in this study consists of a time average over $t$ together with a spatial average over the homogeneous $x$ and $z$ directions. This operation commutes with addition, multiplication by a scalar, differentiation and integration. In the notation adopted here, the Reynolds average is denoted by an overbar. Any variable $\phi$ can be decomposed into a mean field $\overline{\phi}$ and a fluctuating field $\phi'$, satisfying the following rules:

$$\phi = \overline{\phi} + \phi' \tag{5}$$

$$\overline{\phi} = \overline{\overline{\phi}}, \overline{\phi'} = 0 \tag{6}$$

Applying the Reynolds-averaging operation, the Reynolds-averaged form of (1) and (2) is obtained:

$$\boldsymbol{\nabla} \cdot \overline{\boldsymbol{u}} = 0 \tag{7}$$

$$\partial_t \overline{\boldsymbol{u}} + (\overline{\boldsymbol{u}} \cdot \boldsymbol{\nabla})\overline{\boldsymbol{u}} = -\overline{\boldsymbol{\nabla} p} + \nu \nabla^2 \overline{\boldsymbol{u}} + 2\Omega \boldsymbol{k} \times \overline{\boldsymbol{u}} + \nabla \boldsymbol{\tau}_{Re} \tag{8}$$

Where $\overline{\boldsymbol{u}} = (\overline{u}, \overline{v}, \overline{w})^{\mathrm{T}}$ is the averaged velocity, $\boldsymbol{\tau}_{Re} = -\overline{\boldsymbol{u}' \otimes \boldsymbol{u}'}$ is the Reynolds shear stress tensor.

To obtain high-fidelity reference data, equations (1) and (2) are solved numerically using Nek5000, an open-source code based on the spectral-element method. The friction Reynolds number is set to $Re_\tau = 180$. The computational domain has dimensions $L_x : L_y : L_z = 4\pi h : 2h : 2\pi h$ in the streamwise ($x$), wall-normal ($y$) and spanwise ($z$) directions, respectively. Time integration is performed using a third-order backward differentiation scheme (BDF3) with a maximum Courant number of approximately 0.5; and the sampling period satisfies $T u_\tau / h > 40$. The spatial discretization is employed seventh-order polynomials, with 20, 12 and 16 spectral elements in the $x$, $y$ and $z$ directions, respectively. The wall-normal element is stretched from the uniform distribution according to the hyperbolic tangent distribution:

$$y/h = 1 + \frac{\tanh(2.4y'/h - 2.4)}{\tanh(2.4)} \tag{9}$$

where $y'$denotes the positions of the spectral-element edges, uniformly distributed over $[0,2h]$ ( the nodes are located at the Gauss–Lobatto–Legendre quadrature points within each mapped element). The spatial and temporal resolution satisfies the grid-independence requirement, and the statistical averaging period is sufficiently long. The computed profiles of $\overline{u}$ agrees with those of Kim [22] to within 2%, as shown in Fig. 1.

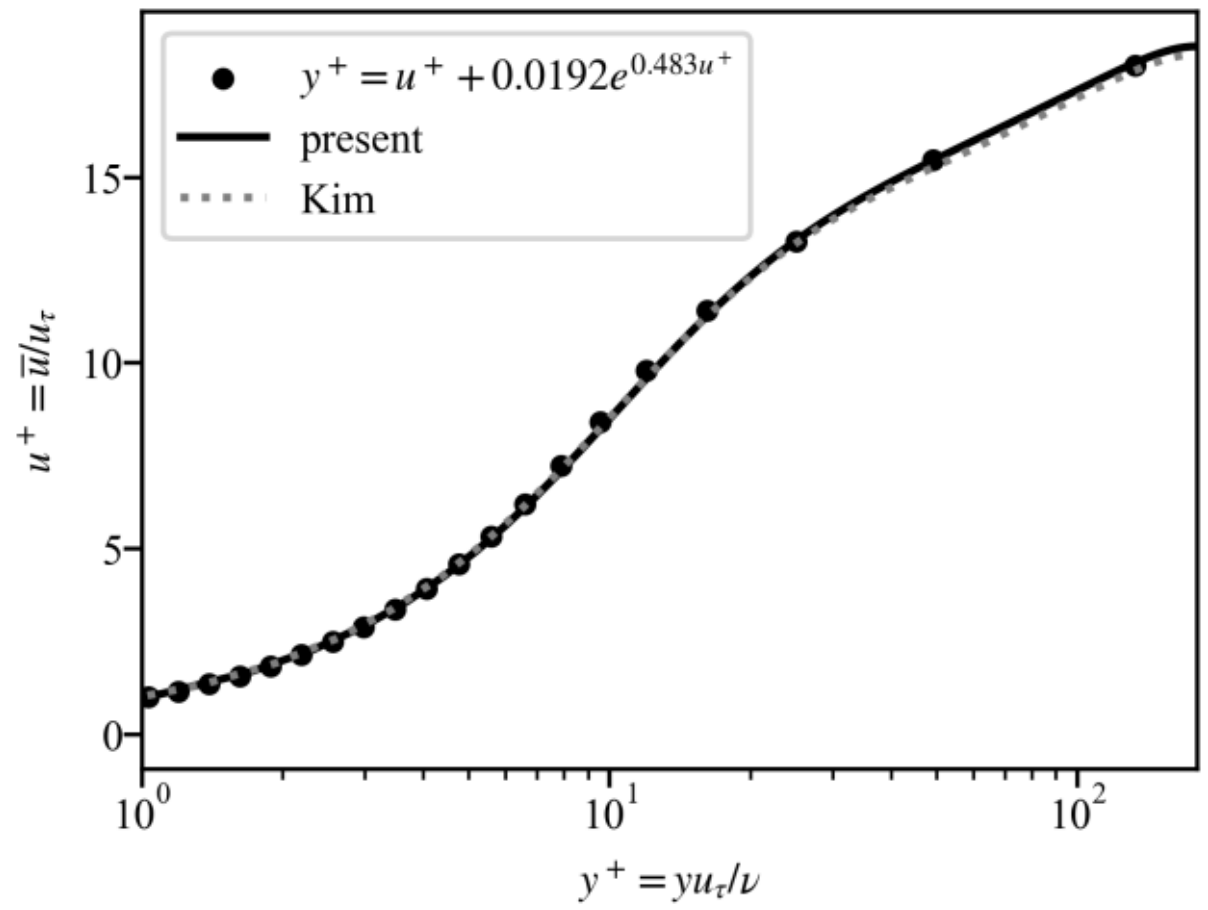

Fig. 1. Comparison of the present DNS results with those of Kim [22] and the theoretical solution.

The parameters adopted in the present work are listed in table I. $Re_b$ is the Reynolds number based on the bulk velocity, and $Re_{\tau,\mathrm{L}}$ and $Re_{\tau,\mathrm{T}}$ are the friction Reynolds numbers at the leading and trailing sides, respectively.

TABLE I. CALCULATION PARAMETERS.

| Case | $Ro_\tau$ | $Re_b$ | $Re_{\tau,\mathrm{L}}$ | $Re_{\tau,\mathrm{T}}$ |
|---|---|---|---|---|
| ST | 0 | 2850 | 180 | 180 |
| A1 | 0.1 | 2840 | 177 | 183 |
| A3 | 0.3 | 2880 | 167 | 191 |
| A5 | 0.5 | 2980 | 154 | 202 |
| A7 | 0.7 | 3000 | 148 | 207 |
| A9 | 0.9 | 3020 | 144 | 209 |
| B1 | 1 | 3020 | 142 | 210 |
| B3 | 3 | 2890 | 134 | 217 |
| B5 | 5 | 2880 | 130 | 218 |
| B7 | 7 | 2830 | 129 | 219 |
| B9 | 9 | 3020 | 130 | 218 |
| C1 | 10 | 3080 | 130 | 218 |
| C3 | 15 | 3390 | 133 | 215 |
| C5 | 20 | 3720 | 138 | 206 |
| A2 | 0.2 | 2860 | 172 | 187 |
| A8 | 0.8 | 3010 | 146 | 208 |
| B2 | 2 | 2950 | 133 | 216 |
| B8 | 8 | 2970 | 129 | 219 |
| C2 | 12 | 3200 | 131 | 216 |
| C4 | 17 | 3220 | 135 | 212 |

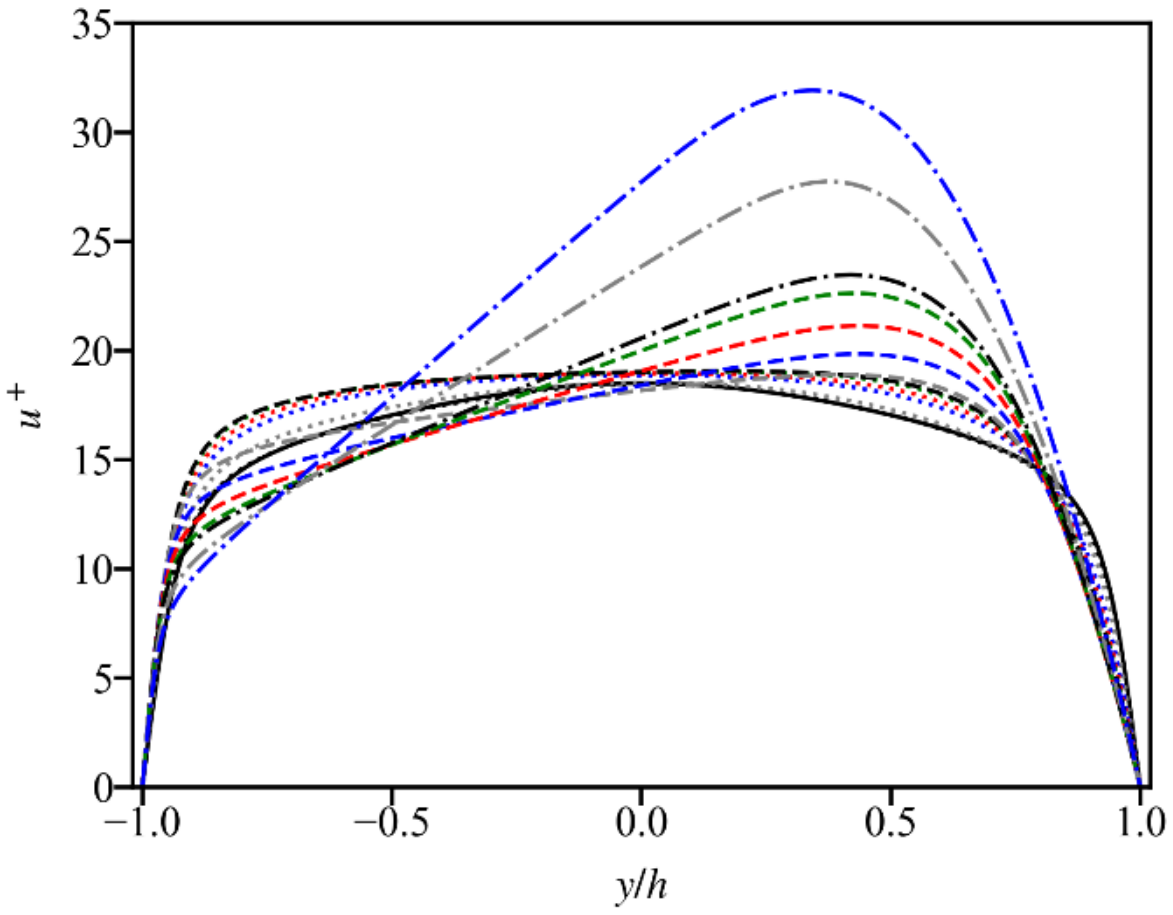

Fig. 2. Averaged streamwise velocity profiles for the different cases.

Fig. 2 shows that $u^+ = \overline{u}/u_\tau$ departs from the non-rotating case only when $Ro_\tau > 0.5$; the velocity profile near the trailing side is blunt, whereas that near the leading side exhibits a deficit. Because the velocity profiles on the two walls are not similar, simple analytical expressions can not describe this difference, and $\overline{u}$ must be calculated from iterated computation. Extracting the $y$ component of (8) and keeping the non-zero terms by streamwise and spanwise homogeneities:

$$\frac{\mathrm{d}\overline{u'v'}}{\mathrm{d}y} = -\frac{\overline{\partial p}}{\partial x} + \nu\frac{\mathrm{d}^2\overline{u}}{\mathrm{d}y^2} \tag{10}$$

$$\frac{\mathrm{d}\overline{v'v'}}{\mathrm{d}y} = -\frac{\overline{\partial p}}{\partial y} + 2\Omega\overline{u} \tag{11}$$

Applying the no-slip and no-penetration boundary conditions at the walls and integrating (10) across the full channel height yields the streamwise pressure gradient $-\overline{\partial p/\partial x} = \tau_\mathrm{w}/h$. Under rotating conditions, the leading and trailing sides each possess a local wall shear stress. To distinguish them, we introduce the subscripts 'L' and 'T' when referring specifically to the leading and trailing sides, respectively; a subscript-free quantity denotes the average shear stress or shear velocity. The momentum conservation relation is given by:

$$\tau_\mathrm{w} = \tau_{\mathrm{w,L}} + \tau_{\mathrm{w,S}}, \quad u_\tau = \sqrt{u_{\tau,\mathrm{L}}^2 + u_{\tau,\mathrm{S}}^2} \tag{12}$$

For convenience, when discussing the trailing side the wall-normal distance $y$ is retained in its original form, whereas for the leading side the real wall distance is $2h - y$. For brevity, when discussing each side separately, the wall distances of both sides are nominally denoted by $y$.

A notable outcome is that the streamwise direction (10) is not explicitly affected by rotation; the influence of the Coriolis force manifests indirectly through the $y$-normal Reynolds stress (11). The present study does not pursue the detailed mechanism by which the Coriolis force couples the streamwise and wall-normal turbulent fluctuations, but focuses instead on how rotation modifies the averaged velocity. Equations (10) and (11) contain unclosed Reynolds-stress terms, and the number of unknowns still exceeds the number of equations. Fortunately, DNS (Direct Numerical Simulation) provides an a priori solution of (3) and (4), from which every component of $\boldsymbol{\tau}_{Re}$ can be obtained.

In engineering practice, the eddy-viscosity hypothesis is widely employed to close the Reynolds-stress tensor. For channel flow, this admits a particularly compact formulation. Introducing the eddy viscosity $\nu_\mathrm{t}$ yields a Reynolds stress expression analogous to Newton's law of viscosity:

$$-\overline{u'v'} = \nu_\mathrm{t}\frac{\mathrm{d}\overline{u}}{\mathrm{d}y} \tag{13}$$

From the high-fidelity DNS data, $\overline{u'v}$ can be calculated statistically, from which the eddy viscosity is obtained as:

$$\frac{\nu_\mathrm{t}}{\nu} = \frac{-\overline{u'v'}}{\nu\mathrm{d}\overline{u}/\mathrm{d}y} \tag{14}$$

Rearranging the right-hand side of (14) and introducing the dimensionless quantities: $y^+ = yu_\tau/\nu$ and $u^+ = \overline{u}/u_\tau$, there is

$$\frac{\nu_\mathrm{t}}{\nu} = \frac{-\overline{u'v'}/u_\tau^2}{\mathrm{d}u^+/\mathrm{d}y^+} \tag{15}$$

In the non-rotating case, the eddy viscosity is commonly estimated through the mixing length $l^+ = lu_\tau/\nu$:

$$\frac{\nu_\mathrm{t}}{\nu} = l^+\left|\frac{\mathrm{d}u^+}{\mathrm{d}y^+}\right|\frac{\mathrm{d}u^+}{\mathrm{d}y^+} \tag{16}$$

von Kármán invoked the similarity hypothesis to close the mixing length finally:

$$l^+ = \kappa y^+ \tag{17}$$

van Driest proposed a more elaborate damping correction:

$$l^+ = \kappa y^+\left(1 - \exp(-y^+/A^+)\right) \tag{18}$$

Where $\kappa \approx 0.41$, and $A^+ \approx 26$. None of the above formulations explicitly captures the influence of rotation. Wei [3] and Tao [5] introduced rotation-dependent corrections to both $\kappa$ and $A^+$ in (18), but no consistent improvement was obtained. This is unsurprising given that the mixing-length model is itself a heuristic construct and, by its nature, lacks the capacity to reflect rotational effects. Moreover, $\mathrm{d}u^+/\mathrm{d}y^+$ vanishes near the center plane of channel, and the eddy viscosity obtained from (15) can even become negative which would be catastrophic for computational stability. This motivates an alternative approach: if the concept of eddy viscosity is dropped, one may directly obtain the law of $\mathrm{d}u^+/\mathrm{d}y^+$ and recover $\overline{u}$ by integration, with the no-slip boundary condition at the wall, thereby bypassing the need for computationally expensive DNS. The present study therefore targets $\mathrm{d}u^+/\mathrm{d}y^+$ as the modelling quantity. Its behavior provides an indirect measure of how the Coriolis force affects the velocity field.

Given that $\mathrm{d}u^+/\mathrm{d}y^+$ generally does not admit a simple closed form, the mean velocity field lacks a simple analytical expression. Deep neural network obscures physical detail, while constructing models from elementary functions or algebraic–differential equations is exceedingly difficult. Symbolic regression overcomes both difficulties. Therefore, the present study employs it to investigate the distribution of $\mathrm{d}u^+/\mathrm{d}y^+$.

### *B. A brief introduction of symbold regression*

The core idea of SR is to recast formula learning as a search-and-evolution process over mathematical expressions. Rather than prescribing a fixed model structure (e.g. nested functions), one initializes a population of expressions assembled from a predefined set of basic operators (e.g. addition, multiplication, trigonometric functions, exponentials and logarithms, etc.), each encoded as a manipulable tree. The population then undergoes iterative genetic evolution through crossover, combination, mutation and fitness evaluation: the quality of each individual is assessed according to both fitting accuracy and expression complexity, superior individuals are retained, and partial sub-trees undergo crossover or mutation to generate new, improved expressions. Over many generations, the population converges towards an analytical formula that is both accurate and compact, thereby automatically extracting the hidden mathematical relationships within the data and yielding interpretable regression models.

The pipeline of SR is illustrated in Fig. 3. Starting from a finite set of prescribed operators, the evolutionary process proceeds through successive crossover, recombination and constant optimization; the optimal expression is ultimately selected under the constraints of expression complexity and the loss function.

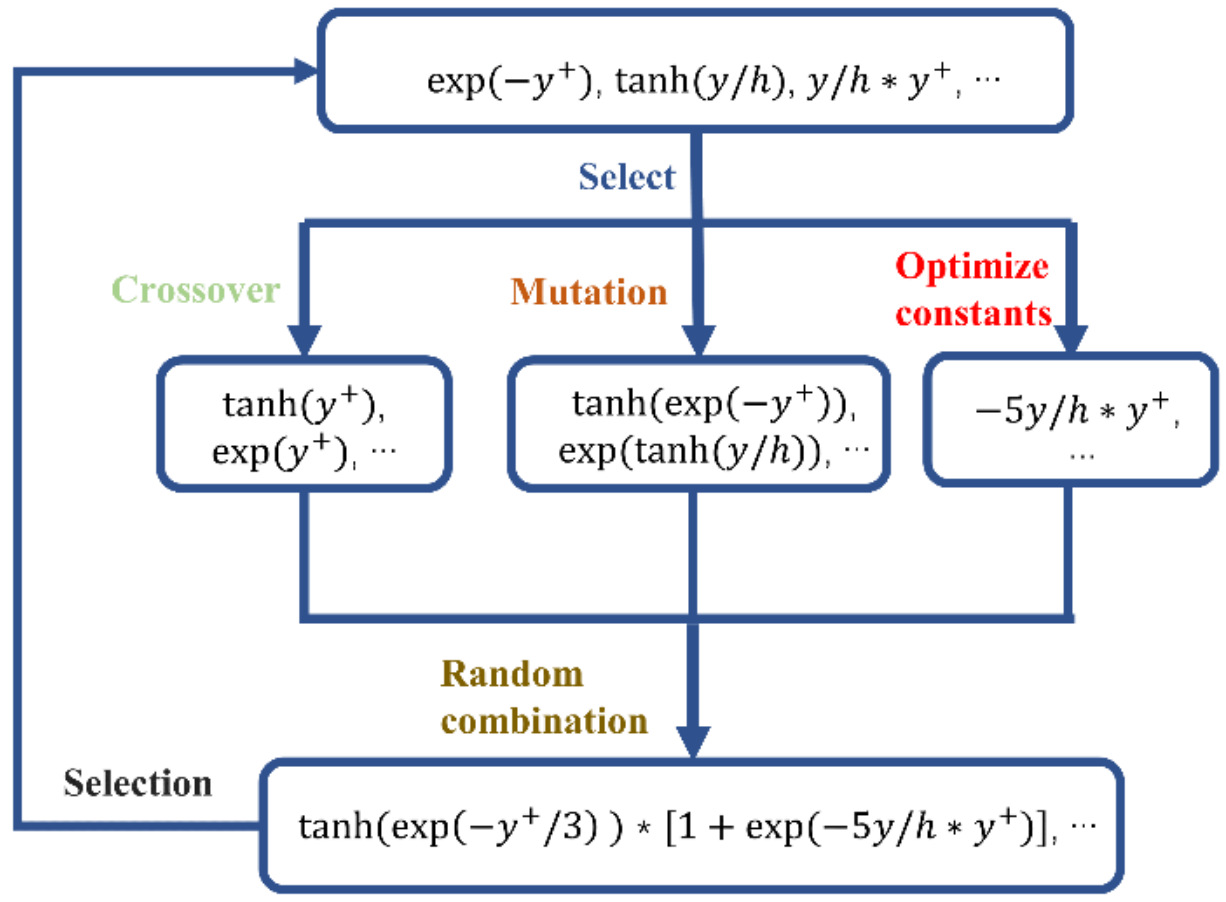

Fig. 3. Schematic of the SR workflow.

To verify the reliability of the symbolic regression, a more compact expression was re-discovered for the law of the wall at $Re_\tau = 180$ under non-rotating conditions, yielding the implicit wall function:

$$y^+ = u^+ + 0.0192e^{0.483u^+} \tag{19}$$

This expression agrees closely with the DNS data and can be viewed as a low-order truncation of Spalding's formula (2). As shown in Fig. 1, the maximum deviation is less than 3%, demonstrating a high level of accuracy. This result confirms that symbolic regression is capable of extracting patterns from small-sample data. Accordingly, the present study employs symbolic regression to investigate how system rotation affects the law of the wall in turbulent boundary layers.

## III. RESULTS AND DISSCUSSIONS

### A. Boundary layer profile

Fig. 4 and 5 show the velocity profiles in semi-logarithmic coordinates. The legend follows that of figure 2, and likewise for all subsequent figures. Comparing the rotating and non-rotating profiles, it is evident that as the rotation number increases from zero, $u^+$ on the leading side is progressively elevated while $u^+$ on the trailing side is progressively suppressed, called the 'Coriolis shift' by Tao [23], and is also captured in laboratory experiments, lending confidence to the present computations. The elevation of the leading-side profile is accompanied by a shift of the velocity peak towards the leading side, implying that, at fixed Reynolds number, the leading-side velocity boundary layer becomes progressively thinner with increasing rotation.

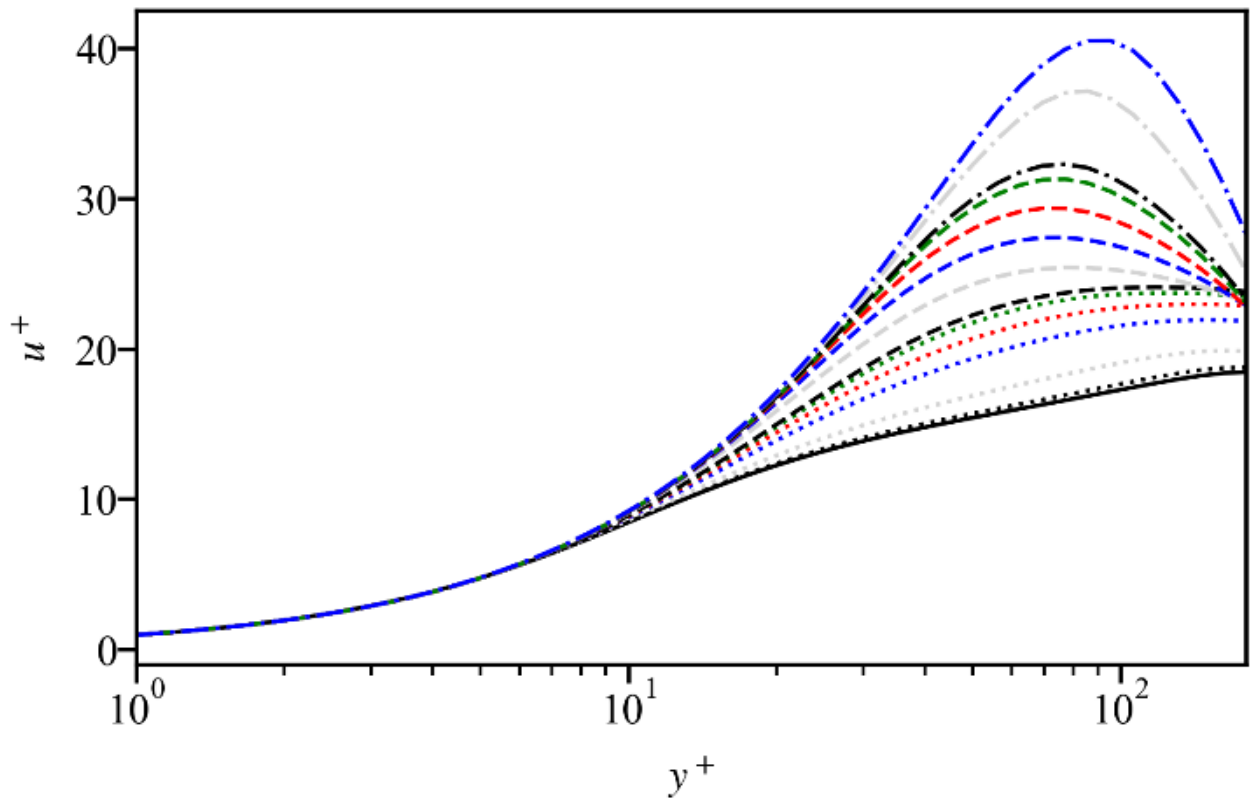

Fig. 4. Dimensionless velocity profiles on the leading side.

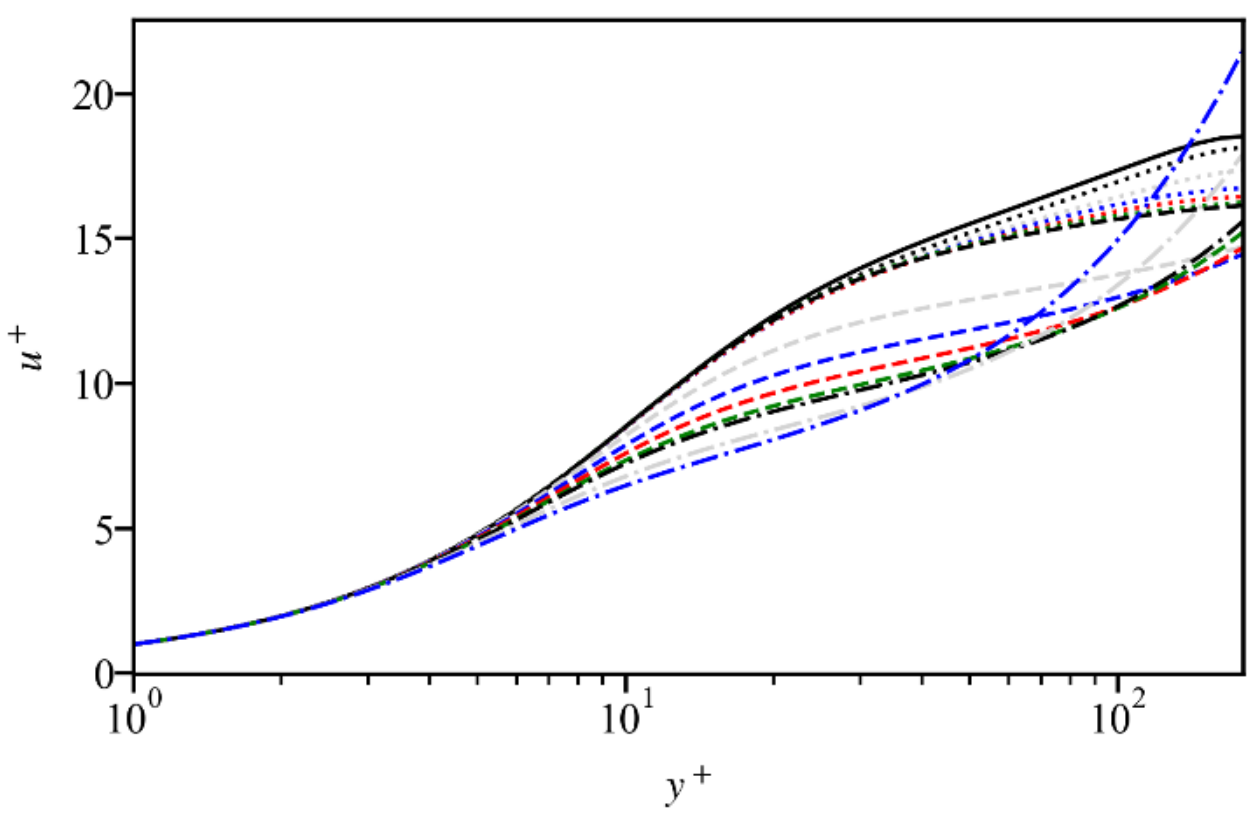

Fig. 5. Dimensionless velocity profiles on the trailing side.

Logarithmic scaling compresses the range of $y^+$ and can lead to underfitting; $\mathrm{d}u^+/\mathrm{d}y^+$ is therefore chosen as the learning target. Reconstructing $u^+$ by integration further reduces random errors. Since the boundary-layer extent on the leading and trailing sides varies with the rotation number, the region $y^+ < 100$ is uniformly adopted as the boundary layer in what follows.

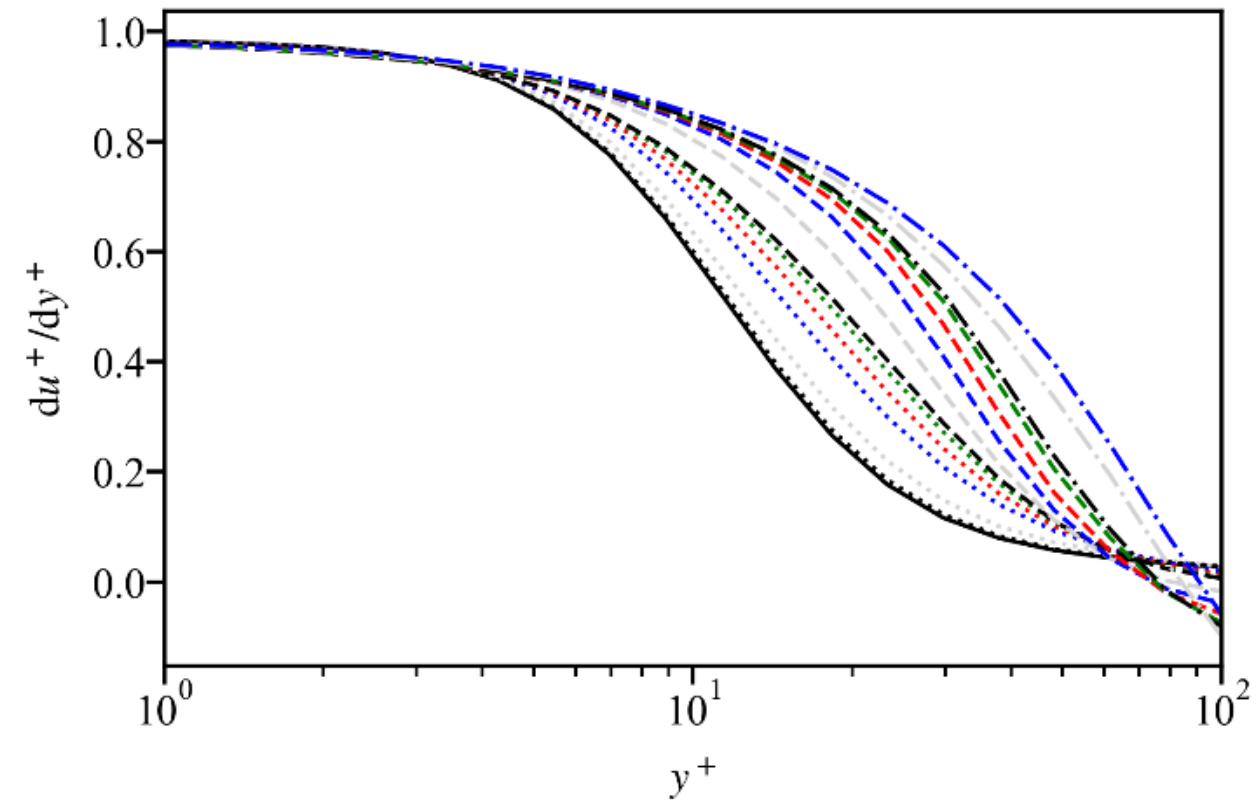

Fig. 6. Dimensionless velocity gradient on the leading side.

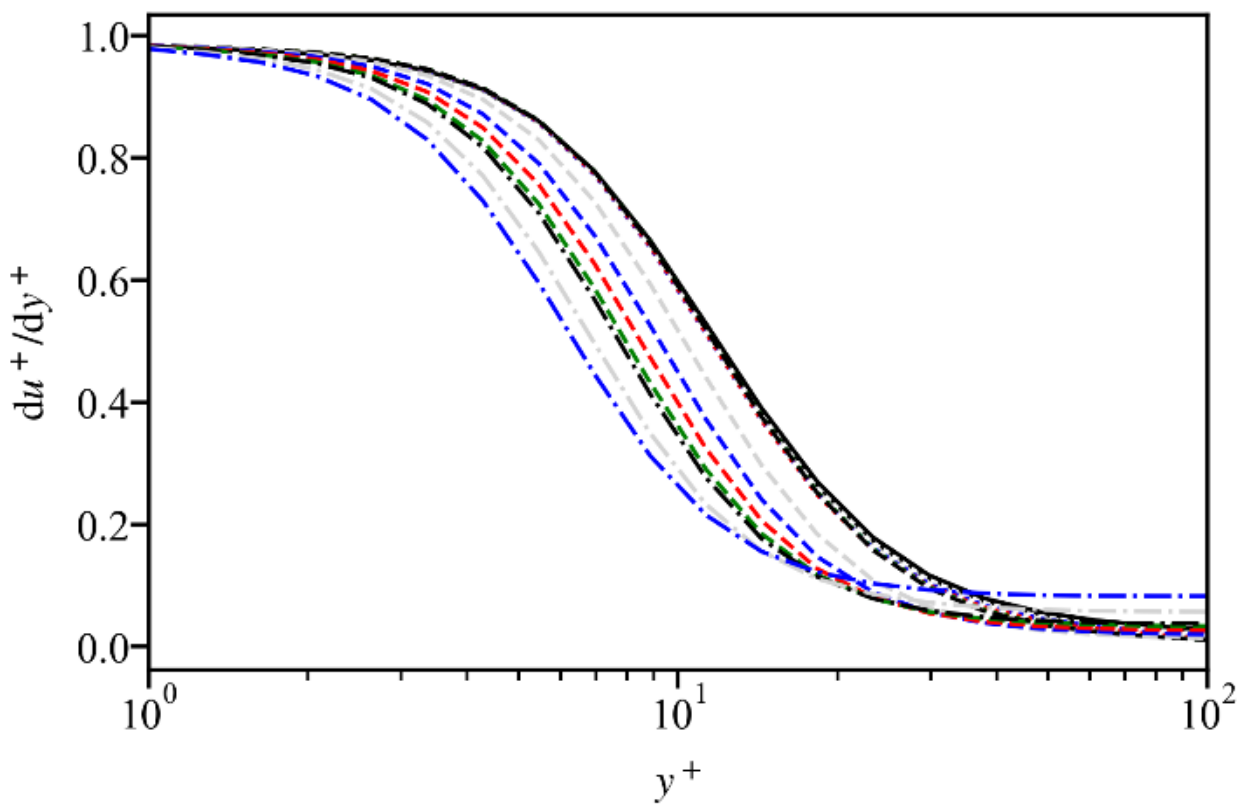

Fig. 7. Dimensionless velocity gradient on the trailing side.

### B. Results of symbol regression

In our study, the PySR package is applied for SR manipulation. Since the leading and trailing sides possess distinct local friction velocities, they are treated separately. The variables and operators employed in the SR are listed in table II:

TABLE II. VARIABLES AND OPERATORS EMPLOYED IN THE SR.

| Variables | $\boldsymbol{y/h, y^+, Ro_\tau}$ |
|---|---|
| Unary operators | $\tanh(x)$, $\exp(x)$, $\ln(x)$ |
| Binary operators | $x+y$, $x*y$, $x/y$, $x^y$ |
| Maximum complexity | 35 |
| Loss function threshold | L1 norm less than 0.01 |

The training terminated after 35 steps, with the L1 norm of loss function reduced below 0.01; the results are shown in Fig. 8. Taking the result at the 30th training step, the expression for the leading side is:

$$\mathrm{d}u/\mathrm{d}y^+ = \left(1 - 1.8\tanh(y/h)\right) * \tanh(16.2Ro_\tau/y^+) * \tanh(Ro_\tau + 0.05h/y) \qquad (20)$$

and that for the trailing side is:

$$\mathrm{d}u/\mathrm{d}y^+ = 0.0046Ro_\tau \tanh(0.41y/hRo_\tau) + \tanh\left(\frac{2.7}{y^+(y/hRo_\tau + 0.024y^+ + 0.038) + 1.08}\right) \quad (21)$$

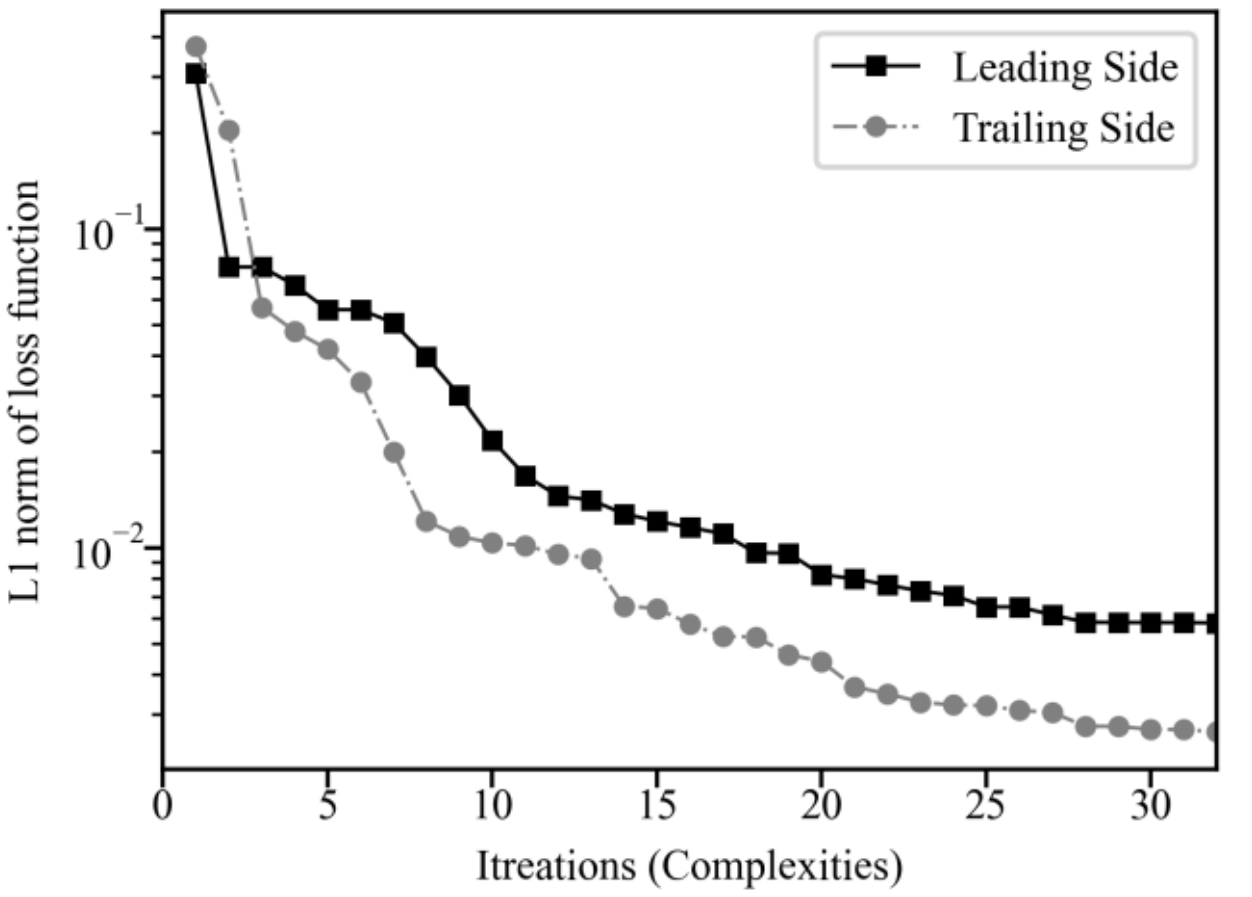

Fig. 8. Evolution of the loss function.

Taking A2, A8, B2, B8, C2 and C4 as the validation set, the predicted $\mathrm{d}u^+/\mathrm{d}y^+$ distributions on the leading side are shown in Fig.9 , where the lines represent the reference data and the symbols the predictions. The maximum error is below 0.02 for $Ro_\tau < 15$ and below 0.05 for $Ro_\tau > 15$. As shown in Fig. 9, the effect of rotation on the leading side is not monotonic: excessively high rotation numbers suppress the turbulent activity on the leading side.

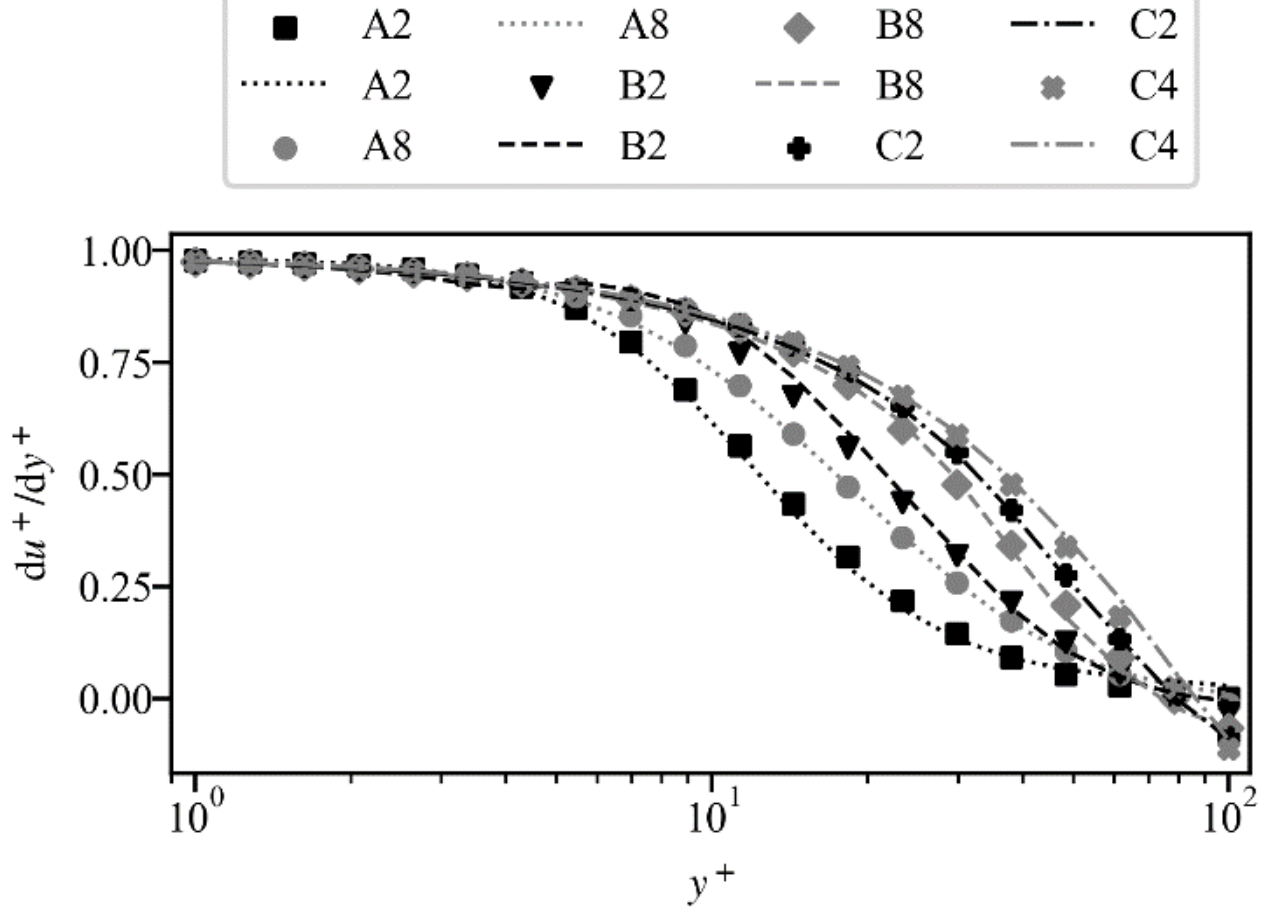

Fig. 9. Predictions of $\mathrm{d}u^+/\mathrm{d}y^+$ for the leading-side test set.

The predictions on the trailing side are superior to those on the leading side. As shown in Fig. 10, the maximum deviation is below 5% in all cases — an appreciable level of accuracy. Integrating $\mathrm{d}u^+/\mathrm{d}y^+$ yields the dimensionless velocity distribution in the boundary layer:

$$u^+(y^+) = \int_0^{y^+} \frac{\mathrm{d}u^+}{\mathrm{d}y^+} dy^+ \quad (22)$$

As shown in Fig. 11 and Fig. 12, the relative error in and below the logarithmic region does not exceed 5% for all cases except the highest rotation number $Ro_\tau = 17$.

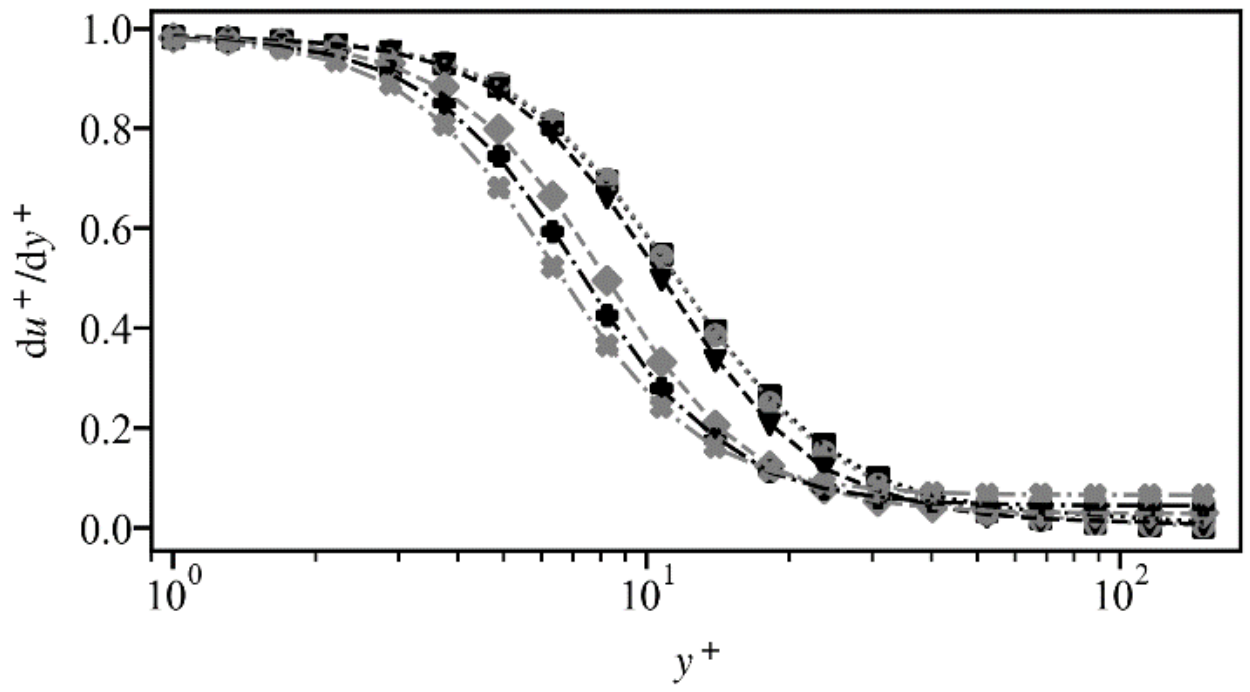

Fig. 10. Predictions of $\mathrm{d}u^+/\mathrm{d}y^+$ for the trailing-side test set.

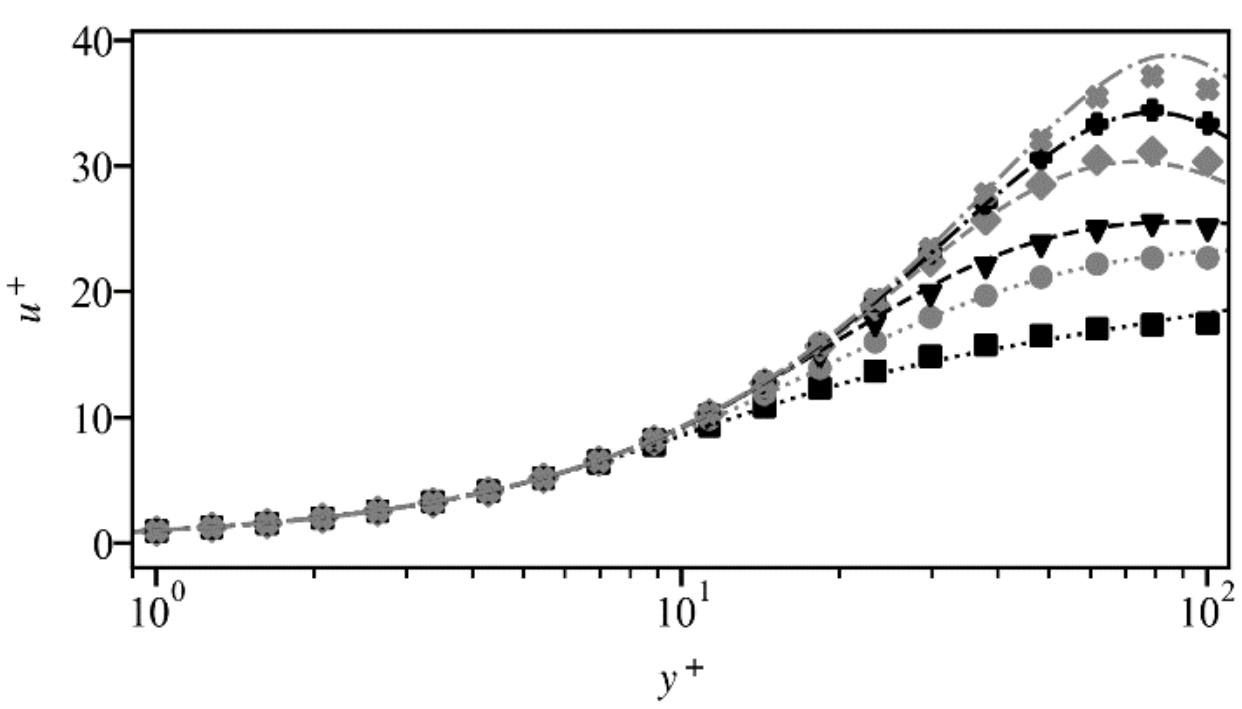

Fig. 11. Integrated results for the leading-side validation set.

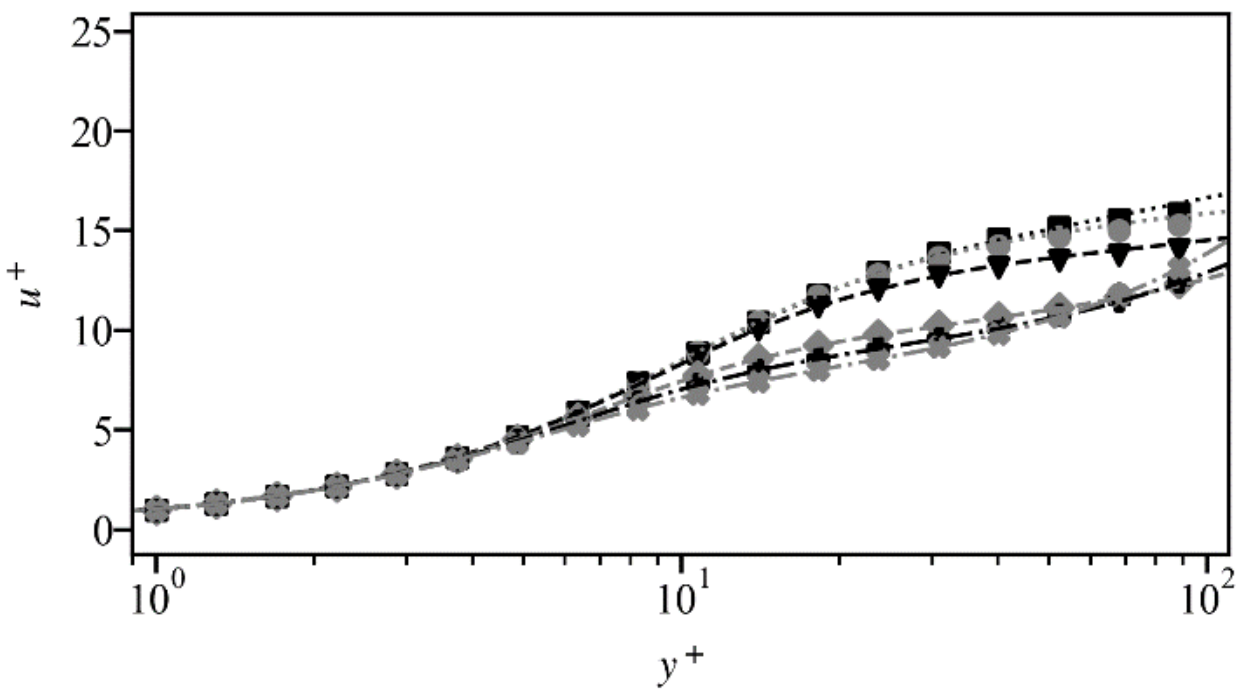

Fig. 12. Integrated results for the trailing-side validation set.

## IV. Conclusion

Symbolic regression was employed in the present work to discover wall-function expressions for turbulent boundary layers at various rotation numbers; the resulting expressions are interpretable, white-box formulations. For rotation numbers below 10, the predictions deviate from the reference data by less than 5%, a high level of accuracy. The wall functions capture how rotation, quantified by the rotation number, shifts the boundary-layer wall function. Incorporating these expressions into RANS or LES forward computations reduces the grid resolution required to resolve the near-wall flow dynamics, and improves the accuracy of wall-bounded turbulent models in turbomachine. Owing to limitations in computational resources, higher-Reynolds-number flows or temperature boundary layers were not investigated in this study, which constitutes a limitation; future work should extend the analysis to a range of Reynolds numbers and cases with heat-transfer.

## ACKNOWLEDGMENT

It is gratefully acknowledged that Beihang University and the National Natural Science Foundation of China (No. 52376042) supported this work.